\documentclass[letterpaper]{article} 
\usepackage{aaai2027}  
\usepackage[hyphens]{url}  
\usepackage{graphicx} 
\graphicspath{{Figures/}}
\usepackage{natbib}  
\usepackage{caption} 
\usepackage{algorithm}
\usepackage{algpseudocode}
\usepackage{newfloat}
\usepackage{listings}
\DeclareCaptionStyle{ruled}{labelfont=normalfont,labelsep=colon,strut=off} 
\floatstyle{ruled}
\newfloat{listing}{tb}{lst}{}
\floatname{listing}{Listing}
\usepackage{booktabs}

\usepackage{amsmath}
\usepackage{xcolor}
\usepackage{pifont}
\usepackage{tikz}
\usetikzlibrary{arrows.meta,positioning,shapes.geometric,fit,backgrounds,calc,decorations.pathreplacing}
\usepackage{placeins}
\usepackage{wasysym}
\usepackage{makecell}
\usepackage{fontawesome5}  

\usepackage{amssymb}

\title{Attack-Resilient Multi-Bit Image Watermarking Framework via Latent Coordinate Sign Clamping}

\author{
    Marjan Beheshti\textsuperscript{\rm 1},
    Ronghua Xu\textsuperscript{\rm 1}
}
\affiliations{
    \textsuperscript{\rm 1}Michigan Technological University\\
    Houghton, MI, USA\\
    mbehes2@mtu.edu, ronghuaxu@mtu.edu
}

\begin{document}

\maketitle

\begin{abstract}
Watermarking is essential for image copyright protection and attribution, yet existing methods suffer from fundamental trade-offs among post-hoc applicability to pre-existing images, multi-bit capacity, training-free deployment, and adversarial resilience.
To address these limitations, we propose a training-free watermarking framework that exploits the overcompleteness of diffusion latent spaces to embed multi-bit payloads into arbitrary images while preserving high visual quality.
Additionally, we develop a three-layer decoding pipeline that combines magnitude-weighted majority voting, BCH, and Chase decoding to maximize message-extraction accuracy in the presence of channel noise.
To counter active adversaries, our framework integrates a steganalysis-resistant noise camouflage layer and a dual-layer forgery defense that combines perceptual hashing with a Reed-Solomon fuzzy commitment scheme over CLIP features.
Empirical evaluations demonstrate that our method applies to both natural and AI-generated images with a scalable $32\text{--}128$-bit capacity.
Furthermore, it effectively mitigates advanced removal vectors ($94.6\%$ decode rate against CtrlRegen+ regeneration attacks) and secures robust forgery detection ($98.8\%$ accuracy via joint CLIP and RS fuzzy commitment mechanism).

\end{abstract}

\section{Introduction}
\label{sec:intro}
While generative AI democratizes high-fidelity content creation, it simultaneously accelerates the spread of misinformation via deepfakes and undermines digital asset attribution \cite{yu2024fake, cao2025secure, nguyen2025survey}.
To restore ecosystem trust and enforce intellectual property protection, digital watermarking has emerged as a vital paradigm for end-to-end provenance verification \cite{sok_watermark, dl_watermark_survey}. As text-to-image models scale, embedding robust, multi-bit watermarks as intrinsic audit trails provides a critical forensic mechanism, enabling downstream defenders to resolve the exact model, user, or platform responsible for an asset's lineage.

Despite significant progress, existing image watermarking methods face a strict four-way trade-off among post-hoc applicability to existing images, multi-bit capacity, training-free deployment, and adversarial resilience.
Post-processing methods~\cite{al2007combined, navas2008dwt, hidden, stegastamp, mbrs, zhang2019robust} directly modify pixel space. While fast and robust to benign transformations, they are easily neutralized by generative regeneration attacks~\cite{zhao23attack}.
Conversely, in-processing techniques~\cite{treering, gaussianshading, prc, xiong2023flexible, optmark} integrate watermarks directly into the content generation pipeline, but they cannot accommodate pre-existing images and require intrusive model modifications or fine-tuning.
Specifically, ZoDiac~\cite{zodiac} leverages latent optimization to bridge the existing-image gap, but it is restricted to zero-bit detection, preventing granular user attribution. Similarly, MetaSeal~\cite{metaseal2026} uses cryptographic watermarks to thwart forgery, yet its reliance on Invertible Neural Networks leaves it critically vulnerable to regeneration attacks. Crucially, no state-of-the-art paradigm simultaneously secures image attribution against both watermark removal and malicious forgery.



\begin{figure}[t]
\centerline{\includegraphics[width=0.48\textwidth]{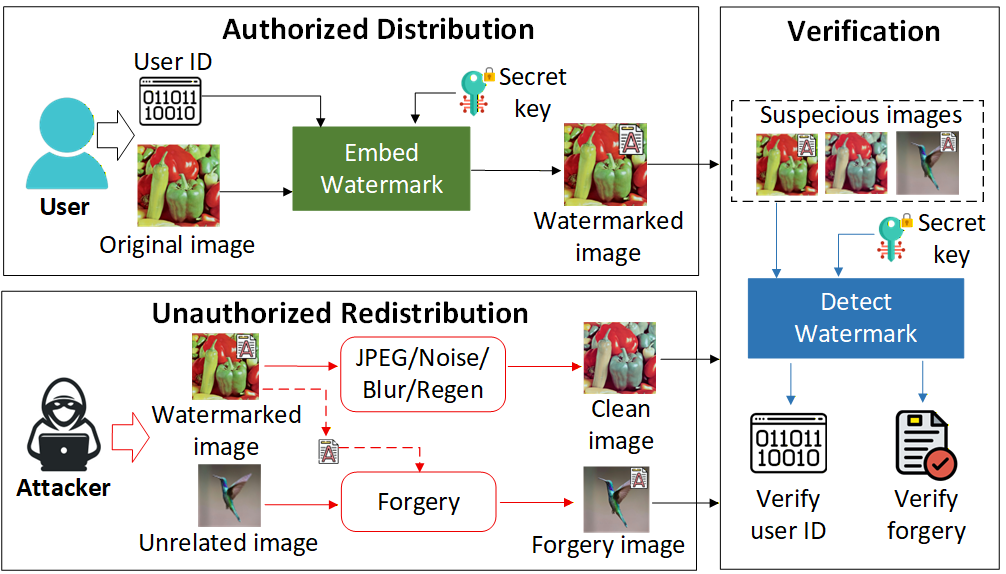}}
\caption{The High-level architecture. \textbf{Left-top:} The platform embeds a user-specific watermark (ID $u$ with secret key $k$) into the original image before distribution. \textbf{Left-bottom:} The image is leaked or redistributed by an unauthorized adversary through diverse attacks. \textbf{Right:} Given suspicious images, the platform detects the watermark to identify the user and verifies image content authenticity.}
\label{fig:system_concept}
\end{figure}

To address these challenges, this paper introduces a training-free, robust, multi-bit watermarking framework tailored for pre-existing imagery.
Figure \ref{fig:system_concept} demonstrates the high-level architecture of our proposed framework.
The watermark embedding pipeline maps the source image to its noisy latent $z_T$ via DDIM (Denoising Diffusion Implicit Models) inversion ~\cite{ddim} and embeds the payload by sign-encoding high-magnitude values into a keyed subset of coordinates, optimized via adaptive $\gamma$-mixing to preserve quality.
We assume an adversary can intercept these watermarked images from the public network to execute watermark removal or forgery attacks.
To counter this, the verification pipeline inverts the suspect image to $z'_T$ and extracts raw bits from the designated coordinates. These are processed through a joint decoder (majority voting, BCH, and Chase decoding) to reconstruct the payload, which is validated against cryptographic content-binding metadata.

\begin{table}[tbp]
\centering
\small
\caption{Comparison of different watermarking methods}
\label{tab:comparison}
\setlength{\tabcolsep}{4pt}
\begin{tabular}{lcccc}
\toprule
\textbf{Method} & 
\makecell[c]{\textbf{Existing} \\ \textbf{images}} & 
\makecell[c]{\textbf{Multi-bit} \\ \textbf{watermark}}& 
\makecell[c]{\textbf{Training-} \\ \textbf{free}}&
\makecell[c]{\textbf{Attack} \\ \textbf{resilience}}\\
\midrule
HiDDeN & \checkmark & \checkmark & $\times$ & $\times$\\
StegaStamp & \checkmark & \checkmark & $\times$ & $\times$\\
Tree-Ring & $\times$ & $\times$ & \checkmark & \LEFTcircle\\
PRC & $\times$ & \checkmark & \checkmark & \LEFTcircle\\
Xiong et al. & $\times$ & \checkmark & $\times$ & \LEFTcircle\\
OptMark & $\times$ & \checkmark & \checkmark & \LEFTcircle\\
ZoDiac & \checkmark & $\times$ & \checkmark & \LEFTcircle \\
MetaSeal & \checkmark & \checkmark & $\times$ & \LEFTcircle \\
\textbf{Ours} & \checkmark & \checkmark & \checkmark & \checkmark \\
\bottomrule
\end{tabular}
{\footnotesize
$\checkmark$\,=\,Supported\quad
$\LEFTcircle$\,=\,Partial\quad
$\times$\,=\,Not supported
}
\end{table}

As illustrated by the comparative analysis in Table \ref{tab:comparison}, our framework introduces a unified architecture that successfully reconciles the trade-offs inherent in competing watermarking paradigms through {four} core innovations.
First, by utilizing DDIM inversion to remap the latent vector and iteratively optimize the initial noise map of a pre-trained, frozen LDM (Latent Diffusion Model), we achieve an entirely training-free deployment that routinely outperforms traditional pre-training and fine-tuning baselines.
Second, the framework is post-hoc applicable. It embeds multi-bit payloads into the latent space of arbitrary natural or synthetic images.
This expanded capacity allows the watermark to convey the rich metadata required for definitive source attribution.
Third, by operating directly in the latent space rather than altering raw pixel values, the embedded payloads remain perceptually unnoticeable while maintaining resilience to adversarial conditions, most notably by resisting Stable Diffusion-based regeneration attacks.
Finally, our verification paradigm integrates content-binding with a cryptographic commitment framework rather than relying on simple presence detection, thereby systematically mitigating the risk of adversarial watermark forgery and replay attacks.

Our main contributions are:
\begin{itemize}

    \item We propose a training- and fine-tuning-free framework that embeds multi-bit identifiers ($32$--$128$\,bits) into arbitrary pre-existing images. The approach demonstrates robust resilience against severe distortions, achieving a $94.6\%$ exact 32-bit decoding rate under diffusion-based regeneration attacks.
    

    \item We demonstrate that clamping $\sim 2\%$ of diffusion latent dimensions ($16{,}384$ for a $512{\times}512$ image) structurally protects the watermark without disrupting reconstruction quality $\text{SSIM} \approx 0.73$). We also expose a non-monotonic interplay between encoding amplitude and $\gamma$-mixing, establishing an empirical operating threshold ($A = 3.0$) that balances signal gain and distortion.


    \item We develop a three-layer decoding pipeline (magnitude-weighted majority voting, BCH, and Chase decoding~\cite{chase1972}) that boosts extraction accuracy from $81.8\%$ to $91.7\%$ over noisy diffusion channels. This architecture is coupled with a steganalysis-resistant noise camouflage layer (mean KS $p = 0.526$) and a dual-layer forgery defense that combines perceptual hashing with a cryptographic fuzzy commitment scheme over CLIP~\cite{clip} features, successfully thwarting up to $98.8\%$ of watermark transplantation attacks.
    
\end{itemize}

\begin{figure*}[t]
\centering
\resizebox{\textwidth}{!}{%
\includegraphics[width=\textwidth]{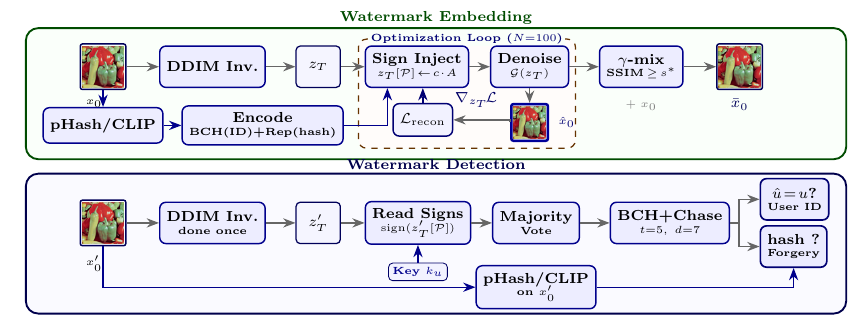}}
\caption{Overview of the embedding (top) and detection (bottom) pipelines. During embedding, the input image is DDIM-inverted, watermark signs are injected at secret with amplitude $A$, and $z_T$ is iteratively optimized to preserve image quality. The watermarked image is $\gamma$-mixed with the original to achieve the target SSIM. During detection, DDIM inversion is performed once; each candidate user ID requires only unscrambling, majority voting, and BCH decoding. The diffusion model remains frozen throughout.}
\label{fig:pipeline}
\end{figure*}

\section{Related Work}
\label{sec:related}

\subsection{Latent-Space Watermarking Methods}

{Latent-space watermarking preserves image fidelity by 
embedding payloads directly into the intermediate 
representations or initial noise of generative models. 
Tree-Ring~\cite{treering} embeds Fourier-domain patterns 
in the initial noise for zero-bit detection via DDIM 
inversion, and RingID~\cite{ci2024ringid} extends this 
to multi-key identification by assigning distinct ring 
patterns to different users. Gaussian 
Shading~\cite{gaussianshading} utilizes key-dependent 
Gaussian noise patterns. However, these methods embed 
watermarks only during generation and cannot be applied 
to pre-existing images. ZoDiac~\cite{zodiac} bridges this 
gap by applying static concentric Fourier rings to 
pre-existing images via latent optimization, but remains 
a zero-bit scheme incapable of user attribution. }
To embed multi-bit payloads, in-processing methods like PRC~\cite{prc}, Xiong~et~al.~\cite{xiong2023flexible}, and OptMark~\cite{optmark} inject tracking data via pseudorandom sign-encoding or multi-step latent perturbations. Although resilient to regeneration attack, they cannot support pre-existing images and require model fine-tuning or architectural modifications. Similarly, latent-embedding schemes like DistSeal~\cite{rebuffi2026learning} and RAIN~\cite{sun2026rain} avoid structural changes but necessitate intensive post-hoc training to guarantee encoder-decoder consistency and reliable watermark recovery.

Our framework bridges this architectural gap by porting the sign-encoding paradigm of PRC into ZoDiac's image optimization pipeline without training an encoder-decoder pair.
Crucially, we substitute ZoDiac's static ring geometry with user-specific signed coordinates in the spatial domain of $z_T$. To optimize error correction over the diffusion channel, we deploy a concatenated BCH code paired with Chase decoding~\cite{chase1972} rather than PRC’s traditional LDPC architecture (See Section~\ref{sec:method}).

\subsection{Forgery Defense and Content Binding}
\label{sec:forgery_defense}
Latent-space watermarking is highly susceptible to forgery (replay) attacks, wherein adversaries transfer static watermark signatures (e.g, Tree-Ring and Gaussian Shading) onto unwatermarked assets without knowledge of the underlying message~\cite{muller2025forgery, jain2025forgery}.
Specifically, ~\citet{muller2025forgery} utilize cross-architecture latent manipulation to imprint target watermarks, while Jain et al.~\cite{jain2025forgery} exploit the many-to-one pixel-to-latent mapping to craft black-box perturbations from a single reference image.
Beyond fixed-pattern exploits, sophisticated adversaries can bypass naive protection by developing adversarial hash collisions between visually disparate media~\cite{neuralhash_break}. 
To counter this, content-dependent watermarking binds payloads directly to image semantics. While frameworks like MetaSeal~\cite{metaseal2026} enforce semantic cryptographic signatures, they require resource-intensive encoder-decoder training.
In contrast, our framework introduces a training-free content-binding mechanism: we jointly encode a perceptual hash with the user ID and apply a cryptographic fuzzy commitment scheme~\cite{juels1999fuzzy} over CLIP features to secure the authentication pipeline against adversarial collision vectors.
\section{Methodology}
\label{sec:method}

\subsection{Threat Model}
\paragraph{Assumptions} Adhering to Kerckhoffs's principle, we assume the adversary possesses comprehensive knowledge of the watermarking framework—including embedding and detection procedures—and access to watermarked samples, while the user's private cryptographic key remains strictly confidential.

\paragraph{Adversary Models} We evaluate two distinct adversarial objectives:
(1)~\emph{Removal attackers} aim to degrade or erase the embedded watermark to prevent user-ID recovery. They may deploy standard transformations (JPEG, noise, blur, contrast adjustments), neural image compression~\cite{cheng20}, or diffusion-based regeneration attacks~\cite{ctrlregen}.
(2)~\emph{Forgery attackers} attempt to transplant a watermark from a source asset into an unrelated target image, exploiting the fixed-pattern vulnerability of zero-bit schemes to trigger false provenance attribution.

\paragraph{Protector Objectives} The protector (content owner or hosting platform) maintains white-box access to the diffusion model and holds the secret cryptographic keys for all registered users. The protector embeds a user-specific signature prior to image distribution. Upon auditing a suspect image, the detector extracts the payload to fulfill a dual objective: (1)~\emph{attribution} (reliably identifying the distributing user) and (2)~\emph{forgery detection} (determining whether the watermark was maliciously transplanted).

\subsection{Overview}


Given a source image $x_0$ and a user identifier $u$, our framework generates a watermarked asset $\bar{x}_0$ that preserves structural fidelity ($\text{SSIM} \geq 0.92$) while guaranteeing accurate payload recovery under adversarial degradation. As illustrated in Figure~\ref{fig:pipeline}, the embedding pipeline proceeds through
three core stages: 
(1)~DDIM inversion to extract the initial noisy latent $z_T$, 
(2)~iterative optimization of $z_T$ with amplified sign 
encoding re-injected at cryptographically keyed coordinates 
every iteration, and (3)~$\gamma$-mixing with the source 
image to meet the target SSIM.
The verification pipeline mirrors this trajectory: it applies DDIM inversion to a suspect image, extracts the latent sign bits at the designated coordinates, and reconstructs the user ID via a joint decoding pipeline comprising majority voting, BCH error correction, and Chase decoding.
To fortify the security boundary, the core architecture is augmented with per-user key-based coordinate isolation, noise camouflage to resist statistical steganalysis, and a dual-layer content-binding mechanism (perceptual hashing and cryptographic fuzzy commitment) to thwart forgery attacks.
The primary mathematical notations are summarized in Table 1 from the supplementary material.


\subsection{Embedding Pipeline}
\label{sec:sign_encoding}

Algorithm~\ref{alg:embed} describes the embedding procedure.
The latent vector $z_T$ corresponding to image $x_0$ is obtained via DDIM inversion, denoted as $\mathcal{G}'(E(x_0))$, where $E$ is a pre-trained VAE (Variational Autoencoder) encoder.
The pipeline encodes a user ID into the signs of a small subset of latent coordinates. Each user $u$ holds an independent 256-bit key $k_u$, from which $m = 317$ watermark positions $\mathcal{P} \subset \{1, \ldots, n\}$ are derived via SHA-256.
The 32-bit user ID is BCH(63,36,$t{=}5$)-encoded into 63 coded bits, each repeated 5 times and mapped to $\pm 1$, yielding a codeword $c \in \{-1,+1\}^m$ after per-user pseudorandom scrambling.
Each selected coordinate is then set to $z_{T,p_i} \leftarrow c_i \cdot A$ (amplitude $A = 3.0$), modifying only ${\sim}2\%$ of the $n = 16{,}384$-dimensional latent space.
The VAE decoder $D$ then reconstructs the watermarked image $\hat{x}_0 = D(\mathcal{G}(z_T))$ from the DDIM-denoised latent, prior to $\mathcal{L}_{\text{recon}}$ optimization.

Throughout this optimization process, the weights of the diffusion model remain frozen, while gradients are backpropagated exclusively to update the latent representation $z_T$. 
Similar to ZoDiac~\cite{zodiac}, the reconstruction objective $\mathcal{L}_{\text{recon}}$ is formulated as a joint loss function that combines pixel-level Mean Squared Error (MSE), the Structural Similarity Index Measure (SSIM), and Watson's perceptual loss.
Consequently, the computational overhead of the embedding phase is fundamentally dominated by the $N \times T$ diffusion forward passes required during the optimization loop.
We provide comprehensive details on DDIM inversion and the encoding process in Sections 1 and 4 of the supplementary material.

\begin{algorithm}[!ht]
\caption{Watermark Embedding}
\label{alg:embed}
\begin{algorithmic}[1]
\Require Image $x_0$, key $k_u$, user ID $u$, amplitude $A$, SSIM threshold $s^*$, iterations $N$
\Ensure Watermarked image $\bar{x}_0$
\State $z_T \gets \mathcal{G}'(E(x_0))$
\State $c \gets \textsc{Encode}(k_u, u)$
\Repeat{ $N$ times}
    \State $z_{T,p_j} \gets c_j \cdot A$ for all $p_j \in \mathcal{P}$
    \State $\hat{x}_0 \gets D(\mathcal{G}(z_T))$
    \State Update $z_T$ via $\nabla_{z_T} \mathcal{L}_{\text{recon}}(x_0, \hat{x}_0)$
\Until{$N$ iterations}
\State $z_{T,p_j} \gets c_j \cdot A$ for all $p_j \in \mathcal{P}$ \Comment{final re-injection}
\State $\hat{x}_0 \gets D(\mathcal{G}(z_T))$
\State $\gamma^* \gets \min\, \gamma \ \text{s.t.}\ \text{SSIM}(\hat{x}_0 + \gamma(x_0 - \hat{x}_0),\, x_0) \geq s^*$
\State \Return $\bar{x}_0 \gets \hat{x}_0 + \gamma^*(x_0 - \hat{x}_0)$
\end{algorithmic}
\end{algorithm}

\subsection{Detection Pipeline}

Algorithm~\ref{alg:detect} describes the detection procedure.
The most computationally intensive operation—the DDIM inversion—is executed exactly \emph{once}, independent of the number of candidate user identifiers being evaluated.
For each candidate validation, the procedure merely extracts $m$ values from the already-recovered latent vector $z'_T$, performs a magnitude-weighted majority vote over each group of $r$ repetitions to reconstruct the coded bits, and invokes BCH and Chase decoding.
Compared to the initial inversion, these subsequent evaluation steps incur negligible computational overhead.
As a result, for a database containing $K$ registered users, the overall detection complexity scales as $\mathcal{O}(T_{\text{DDIM}} + K \cdot m)$, where the inversion latency $T_{\text{DDIM}}$ asymptotically dominates.

\paragraph{Chase-BCH Decoding}
Standard BCH decoding fails when more than $t{=}5$ coded bits are wrong. However, the magnitude-weighted majority vote produces not only a hard decision per coded bit but also a \emph{confidence} value---the absolute value of the weighted sum. Low confidence indicates the group of repeats disagreed, suggesting the bit is likely wrong. Chase decoding~\cite{chase1972} exploits this: it identifies the $d$ least-confident coded bits, tries all $2^d$ combinations of flipping them, and attempts BCH decoding on each. If any combination reduces the error count to $\leq t$, decoding succeeds. With $d{=}7$, the decoder tries 128 combinations. In the best case, this corrects up to $t + d = 12$ errors, though the actual gain depends on the correlation between error positions and low-confidence bits.


\begin{algorithm}[!ht]
\caption{Watermark Detection}
\label{alg:detect}
\begin{algorithmic}[1]
\Require Test image $x'_0$, key $k_u$, candidate user ID $u$, message bits $b$
\Ensure Match $\in \{\text{True}, \text{False}\}$
\State $z'_T \gets \mathcal{G}'(E(x'_0))$
\State $v_i \gets z'_{T, p_i}$ for $i = 1, \ldots, m$ \Comment{read $m$ positions}
\State Unscramble $v$ using seed from $k_u$
\For{$j = 1$ to $n_c$}
    \State $w_j \gets \textstyle\sum_{l \in \text{repeats}(j)} v_l$ \Comment{weighted vote}
    \State $\hat{c}_j \gets \mathbf{1}[w_j > 0]$; \quad $\text{conf}_j \gets |w_j|$
\EndFor
\State $\widehat{msg} \gets \textsc{Chase-BCH-Decode}(\hat{c}, \text{conf}, d{=}7)$
\State $\hat{u} \gets \textsc{BitsToID}(\widehat{msg}_1, \ldots, \widehat{msg}_b)$
\State \Return $\hat{u} = u$
\end{algorithmic}
\end{algorithm}

\subsection{Steganalysis Defense: Noise Camouflage}
\label{sec:camouflage}

Amplifying $m$ coordinates to magnitude $A$ alters the statistical distribution of $z_T$. An attacker who can DDIM-invert the watermarked image may detect the $m$ positions by identifying coordinates with unusually high recovered magnitudes.
To counter this, we introduce \emph{noise camouflage}: in addition to the $m$ real watermark positions, we amplify $m_d$ \emph{decoy} positions to the same magnitude $A$ with random signs. The attacker observes $m + m_d$ loud coordinates but cannot determine which $m$ carry the message. The detector, which holds the key, knows exactly which positions are real and ignores the decoys during decoding. Critically, decoys must use the \emph{same amplitude} as real positions; using a lower amplitude allows the attacker to separate them by magnitude.
The steganalysis defense is evaluated in Section 6.4 of the supplementary material.

\subsection{Forgery Defense: Content-Binding}
\label{sec:forgery_detection}

Our multi-bit watermark capacity enables defense against forgery by encoding a \emph{content-bound message}---the user ID plus an image-specific fingerprint---within the same 317 positions. 
The system supports two modes: \emph{tracking mode} (32-bit ID only, maximum robustness) and \emph{anti-forgery mode} (32-bit ID + 16-bit content fingerprint via unequal error protection). 
In anti-forgery mode, the user ID uses BCH(63,36,$t{=}5$) with 3 repeats per coded bit (189 positions), while the 16-bit fingerprint uses 8 repeats per bit (128 positions).
At detection, the decoder recovers both components and recomputes the fingerprint from the test image; a mismatch reveals forgery.

We provide two content fingerprinting options: (1)~a perceptual hash (pHash) for practical detection (441/500 forgeries caught), and (2)~a cryptographic fuzzy 
commitment~\cite{juels1999fuzzy} over CLIP features~\cite{clip} that resists adversarial hash manipulation (98.8\% detection).
The fuzzy commitment 
derives a content-binding key $k_{\text{cb}}$ from a 
Reed--Solomon-protected secret locked to the image's CLIP 
features; forgery causes decoding failure and key mismatch. 
Full details are in Section 5
of the supplementary material.

\FloatBarrier
\section{Experiments}
\label{sec:experiments}

\subsection{Experimental Setup}
\label{sec:setup}

\paragraph{Datasets and models}
We evaluate our framework across three distinct benchmarks, with all assets standardized to a resolution of $512 \times 512$: (1)~MS-COCO val2017~\cite{coco}, comprising 500 real-world photographs unexposed to generative models; (2)~DiffusionDB~\cite{diffusiondb}, consisting of 500 synthetic images generated from open-ended text prompts; and (3)~WikiArt~\cite{wikiart}, containing 500 art-style images synthesized from style-specific prompts. Synthetic generations for both DiffusionDB and WikiArt are executed via Stable Diffusion 2.1.

\paragraph{Implementation Details}
All experiments are implemented in PyTorch and executed on a single NVIDIA RTX 5090 GPU.
Our core setup integrates Stable Diffusion 2.1 configured with 50 DDIM denoising steps, $m = 317$ latent coordinates, encoding amplitude $A = 3.0$ (Section~6.1 of the supplementary material),  SSIM quality constraint $s^* = 0.92$, and $N = 100$ optimization iterations. Users are assigned independent 256-bit cryptographic keys $k_u$. In the primary evaluation, a 32-bit user identifier is protected via a concatenated channel coding pipeline consisting of a BCH(63,36,$t{=}5$) outer code and a repetition-5 inner code.
To demonstrate high-capacity, we also evaluate a 128-bit payload mapping to 1,024 latent coordinates protected by a BCH(255,131,$t{=}18$) and repetition-4 coding scheme.


\paragraph{Evaluation Metrics}
To evaluate visual fidelity and distributional generation quality, we assess performance using Peak Signal-to-Noise Ratio (PSNR), the Structural Similarity Index Measure (SSIM), the Fréchet Inception Distance (FID), and CLIP score.
To quantify extraction robustness, we report both \emph{Bit Accuracy} (the proportion of payload bits correctly decoded) and \emph{Decode Rate} (the percentage of images for which the exact 32-bit user identifier is fully recovered).
We adopt decode rate as the primary metric for robustness evaluation under attacks, as it is exponentially harder than bit-level accuracy (See discussion in Section~7.1 of the supplementary material). For forgery detection, we measure the Hamming distance between the decoded perceptual hash and the target image's recomputed hash; a distance exceeding 4 out of 16 bits indicates a detected forgery.

\subsection{Performance Evaluation}

\begin{table*}[tbp]
\centering
\small
\caption{Evaluation of image visual quality, decode rate, and bit accuracy across three benchmarks. Image quality metrics are averaged under the baseline (no-attack) condition. ``Combined'' is the average across 1,500 images (500 per benchmark).}

\label{tab:main_results}
\setlength{\tabcolsep}{5pt}
\begin{tabular}{l cccc ccccc ccccc}
\toprule
& \multicolumn{4}{c}{\textbf{Quality}} & \multicolumn{5}{c}{\textbf{Decode Rate (\%) $\uparrow$}} & \multicolumn{5}{c}{\textbf{Bit Accuracy (\%) $\uparrow$}} \\
\cmidrule(lr){2-5} \cmidrule(lr){6-10} \cmidrule(lr){11-15}
\textbf{Dataset} & SSIM $\uparrow$ & PSNR $\uparrow$ & FID $\downarrow$ & CLIP $\uparrow$ & None & JPEG & Noise & Blur & Ch20 & None & JPEG & Noise & Blur & Ch20 \\
\midrule
MS-COCO & 0.919 & 29.2 & 20.8 & 0.983 & 99.6 & 97.4 & 96.4 & 99.6 & 89.4 & 99.9 & 99.1 & 98.7 & 99.9 & 96.4 \\
DiffusionDB & 0.921 & 28.3 & 11.0 & 0.995 & 100 & 99.4 & 98.0 & 100 & 93.4 & 100 & 99.7 & 99.1 & 100 & 98.2 \\
WikiArt & 0.922 & 29.0 & 16.4 & 0.992 & 100 & 98.8 & 98.2 & 100 & 92.2 & 100 & 99.7 & 99.4 & 100 & 97.6 \\
\midrule
\textbf{Combined} & \textbf{0.921} & \textbf{28.8} & \textbf{8.74} & \textbf{0.989} & \textbf{99.9} & \textbf{98.5} & \textbf{97.5} & \textbf{99.9} & \textbf{91.7} & \textbf{100} & \textbf{99.5} & \textbf{99.1} & \textbf{100} & \textbf{97.4} \\
\bottomrule
\end{tabular}
\end{table*}

\label{sec:main_results}
Table~\ref{tab:main_results} reports evaluations of image quality, decode rate, and bit accuracy across 1,500 images (500 per benchmark).
Protected images are evaluated under five conditions from the WAVES benchmark~\cite{waves}: no attack, JPEG (QF=50), additive Gaussian noise ($\sigma = 0.05$), Gaussian blur (kernel 5, $\sigma = 1.0$), and Cheng20~\cite{cheng20} neural compression (Quality 3).
We use ZoDiac's~\cite{zodiac} public attack suite to guarantee absolute baseline parity.
All reported results use the joint three-layer decoder for payload extraction and verification.

\paragraph{Image Quality}

\begin{figure}[tbp]
\centering
\includegraphics[width=\columnwidth]{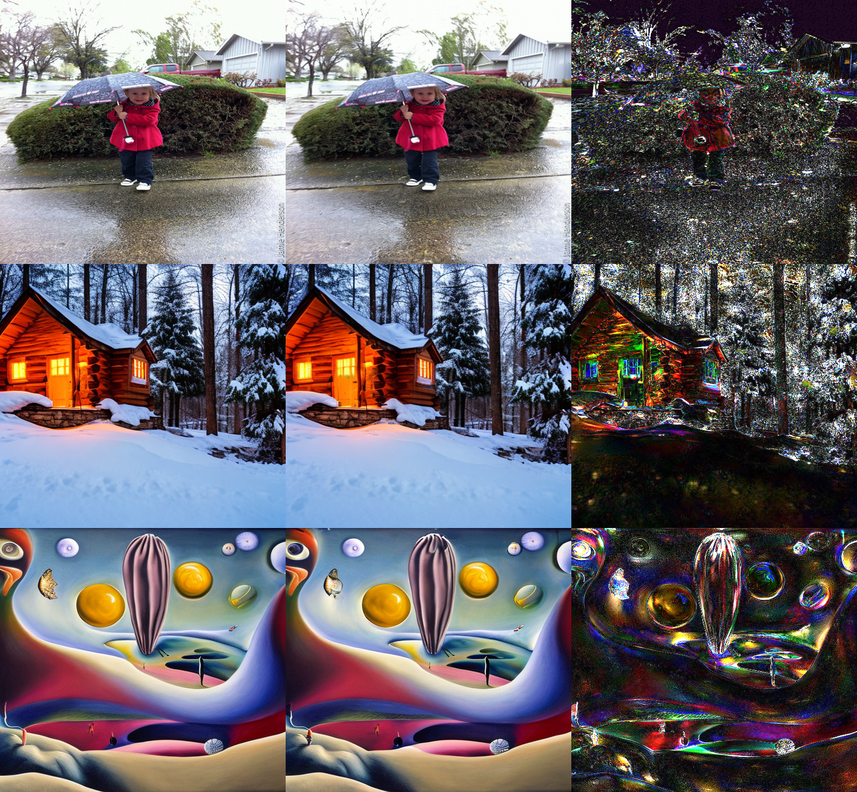}
\caption{Visual quality comparison across three benchmarks (one image per dataset). Left: original. Center: watermarked (SSIM $\geq 0.92$). Right: absolute pixel difference amplified $10\times$. Rows: MS-COCO (top), DiffusionDB (middle), WikiArt (bottom).}
\label{fig:quality}

\end{figure}

Watermarked images preserve high visual fidelity (SSIM = 0.921, PSNR = 28.8 dB, FID = 8.74, CLIP score = 0.989 $\pm$ 0.011), closely aligning with the quality boundaries established by ZoDiac.
Figure~\ref{fig:quality} provides qualitative examples demonstrating that the embedded perturbations remain imperceptible under standard human inspection, with structural residuals becoming apparent only under 10× error amplification.

\paragraph{Watermark Robustness}

The proposed method maintains high decode rate and bit accuracy across all four distortion categories. While Cheng20 neural compression represents the most severe degradation vector, our three-layer decoder achieves a $91.7\%$ decode rate—a significant increase from the $81.8\%$ baseline without Chase decoding. All other configurations exceed a $97\%$ success rate. Chase decoding provides the largest margins of improvement under Cheng20 ($+9.9\%$), successfully reconstructing near-threshold payloads where standard algebraic BCH decoding fails (Section~\ref{sec:ablation}).


\subsection{Attack Resilience}

\begin{table}[tbp]
\centering
\small
\caption{Robustness under CtrlRegen+ regeneration attack (500 MS-COCO images). $\dagger$From original paper. StegaStamp and TreeRing from~\cite{ctrlregen}.}
\label{tab:regen_comparison}
\setlength{\tabcolsep}{5pt}
\begin{tabular}{lccccc}
\toprule
\textbf{Method} & \textbf{Setting} & \textbf{Metric} & \textbf{No Att.} & \textbf{s=0.3} & \textbf{s=0.5} \\
\midrule
\textbf{Ours} & Existing & Decode & 99.6\% & \textbf{94.6\%} & \textbf{55.4\%} \\
ZoDiac & Existing & Det. & 99.8$^\dagger$\% & 18.0\% & 6.0\% \\
StegaStamp & Existing & Det. & 100$^\dagger$\% & ${\sim}$1\% & ${\sim}$0\% \\
TreeRing & Gen. & Det. & 99$^\dagger$\% & ${\sim}$12\% & ${\sim}$0\% \\
\bottomrule
\end{tabular}

\end{table}

\paragraph{Regeneration Attack}


We benchmark against CtrlRegen+~\cite{ctrlregen}, a state-of-the-art diffusion-based removal attack that regenerates imagery via controlled denoising on a surrogate Stable Diffusion 1.5 model.
Table~\ref{tab:regen_comparison} presents robustness comparisons for regeneration across methods.
At the standard quality-preserving step threshold of $t = 0.3$, our method achieves a 94.6\% exact 32-bit decode rate on 500 MS-COCO images. At an aggressive setting of $t = 0.5$, 
the decode rate drops to 55.4\%.

While ZoDiac reports $98.8\%$ detection under the earlier Zhao~et~al. vector~\cite{zhao23attack}, our evaluation demonstrates its accuracy falls to $18.0\%$ against CtrlRegen+ ($t = 0.3$). StegaStamp and Tree-Ring are similarly neutralized. Our framework represents the only post-hoc image watermarking solution that preserves high multi-bit decoding capacity under state-of-the-art regeneration constraints.



\paragraph{Forgery Attack}
To assess resilience against watermark transplantation, we simulate an adversarial forgery protocol: the attacker performs DDIM inversion on a watermarked source image, 
extracts the latent coefficients at the 317 designated watermark coordinates, and implants them into the latent representation of an unrelated target image.
We evaluate two content-fingerprinting modes independently within the same 317 positions.
Table~\ref{tab:fuzzy} reports modification survival rates and forgery detection for both modes.

\textit{(1) Perceptual hash (pHash):} We construct 500 forgery pairs(50 source images $\times$ 10 targets), jointly encoding a 32-bit user identifier and a 16-bit perceptual hash. All 50 authentic images yield a hash Hamming distance of~0 with correct user identification. Under the forgery condition, the mean decoded Hamming distance rises to 7.4/16 (min~2, max~14), with 441/500 (88.2\%) transplanted pairs exceeding the detection 
threshold. The undetected cases correspond to source--target pairs with high visual similarity, producing near-colliding perceptual hashes.

\textit{(2) Cryptographic fuzzy commitment:} We evaluate CLIP-based forgery detection across 1,225 cross-image pairs (50 images, all pairwise combinations). The fuzzy commitment detects 1,210/1,225 (98.8\%) forgeries, 
systematically addressing the hash-collision vulnerability of pHash.



\begin{table}[tbp]
\centering
\small
\caption{Fuzzy commitment: modification survival (500 images) and forgery detection comparison.}
\label{tab:fuzzy}
\setlength{\tabcolsep}{3pt}
\begin{tabular}{lcc}
\toprule
\textbf{Modification} & \textbf{Pass / Total} & \textbf{Rate} \\
\midrule
{Brightness ($\pm 0.1$)} & {500/500} & {100\%} \\
{Gaussian blur ($k{=}5$)} & {500/500} & {100\%} \\
{Noise ($\sigma{=}0.05$)} & {494/500} & {98.8\%} \\
{Noise ($\sigma{=}0.1$)} & {459/500} & {91.8\%} \\
{JPEG (QF=50)} & {481/500} & {96.2\%} \\
{JPEG (QF=25)} & {453/500} & {90.6\%} \\
\midrule
{\textbf{All modifications}} & {\textbf{2887/3000}} & {\textbf{96.2\%}} \\
\bottomrule
\\[4pt]
\toprule
\textbf{Method} & \textbf{Forgery det.} & \textbf{Adv.\ resistant} \\
\midrule
{pHash (16-bit)} & {441/500 (88.2\%)} & \ding{55} \\
{CLIP + RS fuzzy commit} & {1210/1225 (98.8\%)} & \checkmark \\
\bottomrule
\end{tabular}
\end{table}

The CLIP-based fuzzy commitment (Section~5 of the supplementary material) is evaluated on 500 images across six standard modifications.
It achieves a $96.2\%$ overall modification survival rate and increases forgery detection from $88.2\%$ (pHash baseline) to $98.8\%$ ($1{,}210/1{,}225$ cross-image pairs).
Notably, the targeted brightness attack that neutralizes pHash at $\text{SSIM} > 0.94$ fails entirely against our CLIP-based features: across 500 test images, the strongest adversarial optimization only reduces the cross-image CLIP distance to $48/128$, far exceeding the Reed--Solomon correction threshold.
While a $\sim 9\%$ false-rejection rate occurs under heavy JPEG compression (QF=25), this trade-off effectively eliminates the structural adversarial vulnerability of traditional perceptual hashing.

\subsection{Model generalizability.}

\begin{table}[tbp]
\centering
\small
\caption{Performance of our method across different Stable Diffusion variants ({500 MS-COCO images}).}
\label{tab:generalizability}
\setlength{\tabcolsep}{5pt}
\begin{tabular}{lcccccc}
\toprule
\textbf{Model} & \textbf{SSIM} & \textbf{PSNR} & \textbf{FID} & \textbf{CLIP} & \textbf{Decode Rate (\%)} \\
\midrule
SD 2.1 & 0.919 & 29.2 & 20.8 & 0.983 & 89.4--99.6 \\
SD 1.5 & 0.921 & 29.8 & 19.2 & 0.964 & 89.8--99.6 \\
\bottomrule
\end{tabular}
\end{table}


Table \ref{tab:generalizability} shows the performance of our training-free, plug-and-play method, which can be transferred directly to other LDM variants.
We test generalization to SD~1.5 (same UNet architecture, different weights). SD~1.5 achieves comparable results on 500 MS-COCO images (SSIM = 0.921, decode rates 89.8--99.6\% across all attacks).
The method is compatible with any LDM sharing the same VAE architecture.


\subsection{Capacity Analysis}
\label{sec:capacity}

\begin{table}[tbp]
\centering
\small
\caption{Capacity--robustness tradeoff.}
\label{tab:capacity}
\setlength{\tabcolsep}{2.5pt}
\begin{tabular}{l cc ccccc}
\toprule
& \multicolumn{2}{c}{\textbf{Quality}} 
& \multicolumn{5}{c}{\textbf{Decode Rate (\%)}} \\
\cmidrule(lr){2-3} \cmidrule(lr){4-8}
\textbf{Config} & SSIM & PSNR 
& None & JPEG & Noise & Blur & Ch20 \\
\midrule
32-bit (317) & .919 & 29.2 & 99.6 & 97.4 & 96.4 & 99.6 & 89.4 \\
128-bit (1024) & .920 & 28.4 & 99.2 & 98.0 & 97.0 & 99.4 & 92.2 \\
\bottomrule
\end{tabular}
\end{table}


Table~\ref{tab:capacity} demonstrates the capacity--robustness trade-off by evaluating 32-bit and 128-bit user identifiers on 500 MS-COCO images.
The 32-bit configuration maximizes resilience by using a repetition-5 inner code and BCH outer correction to withstand all standard distortions.
The 128-bit setup offers $4\times$ the payload capacity—enabling simultaneous encoding of a user ID, timestamp, and metadata—by utilizing 1,024 latent coordinates ($6.3\%$ of the latent space).
Remarkably, despite this larger payload footprint, robustness and visual fidelity remain nearly indistinguishable from the baseline.
Both variations employ the same embedding pipeline, differing only in error-correcting code parameters.

\subsection{Ablation Studies}
\label{sec:ablation}

\begin{table}[tbp]
\centering
\small
\caption{Decoder design impact on Cheng20 decode rate (hardest attack). All results on 1,500 images with the same embedded watermarks.}
\label{tab:bch_vs_mv}
\setlength{\tabcolsep}{5pt}
\begin{tabular}{l cccc}
\toprule
\textbf{Decoder} & \textbf{COCO} & \textbf{DiffDB} & \textbf{WikiArt} & \textbf{All} \\
\midrule
BCH($t{=}5$), no Chase & 74.8 & 86.2 & 84.4 & 81.8 \\
BCH + Chase($d{=}7$) & \textbf{89.4} & \textbf{93.4} & \textbf{92.2} & \textbf{91.7} \\
\bottomrule
\end{tabular}
\end{table}

\paragraph{Decoder Design Impact: BCH vs.\ Chase}
Table~\ref{tab:bch_vs_mv} shows the impact of Chase decoding on Cheng20, the hardest attack. Without Chase, BCH alone achieves 81.8\% by correcting up to 5 coded-bit errors after majority voting. Adding Chase decoding improves this to 91.7\% by trial-flipping unreliable coded bits before BCH correction, rescuing images with 6--12 coded-bit errors at negligible computational cost.


\subsection{Comparison with Prior Work}
\label{sec:comparison}

\begin{table}[tbp]
\centering
\footnotesize
\caption{Comparison with prior watermarking methods.}

\label{tab:comparison_full}
\setlength{\tabcolsep}{4pt}
\begin{tabular}{l ccccc}
\toprule
& \textbf{Ours} & \textbf{ZoDiac} & \textbf{PRC} & \textbf{Xiong} & \textbf{OptMark} \\
\midrule
SSIM  & 0.921 & 0.92 & --- & 0.93 & --- \\
PSNR  & 28.8 & 29.4 & --- & 32.5 & --- \\
\midrule
& \multicolumn{5}{c}{\textit{Decode / Bit acc.\ (ours) vs.\ reported metric}} \\
\midrule
JPEG & 98.5/99.5 & 99.2 & 1.00 & 99.4 & 99.3 \\
G-noise & 97.5/99.1 & 99.6 & 0.94 & 98.0 & --- \\
G-blur & 99.9/100 & 99.6 & --- & 99.6 & --- \\
Ch20/VAE & 91.7/97.4 & 98.6 & --- & --- & 90.4 \\
\midrule
Metric & Dec./Bit & Det. & TPR & Bit acc. & Bit acc. \\
\bottomrule
\end{tabular}

\end{table}


Table~\ref{tab:comparison_full} benchmarks our method against ZoDiac, PRC, Xiong~et~al., and OptMark across quality and robustness. Cross-method comparisons require care due to metric heterogeneity: ZoDiac reports zero-bit detection, PRC evaluates TPR at a fixed FPR, Xiong~et~al.\ and OptMark measure bit accuracy, and we report exact 32-bit decode rates. Exact payload decoding is strictly more challenging than detection or bit-level accuracy.
Despite optimizing for this rigorous metric, our framework remains highly competitive.
Our Cheng20 decode rate ($91.7\%$) is numerically lower than ZoDiac's $98.6\%$, but ZoDiac measures the presence of zero bits—a simpler task that requires only $\sim 60\%$ sign alignment rather than full 32-bit recovery (Section~7 of the supplementary material).
Among multi-bit baselines, our performance matches or exceeds both Xiong~et~al.\ and OptMark across all attacks (Table~\ref{tab:comparison_full}), while representing the only training-free solution applicable to existing images.
PRC occupies a distinct generation-only paradigm with bounds on cryptographic undetectability, making it less directly comparable.
Lastly,
our lower PSNR (28.8\,dB vs.\ 32.5\,dB for Xiong~et~al.) 
represents the inherent cost of post-hoc asset modification, and our high SSIM ($0.921$) demonstrates that this geometric variance is perceptually minor.


\section{Conclusion}
\label{sec:conclusion}
In this paper, we presented a novel watermarking framework that embeds multi-bit payloads into the latent space of arbitrary, pre-existing images without requiring any fine-tuning or retraining of the underlying generative architecture.
We systematically demonstrated that the structural overcompleteness of diffusion latent spaces provides robust multi-bit watermarking under both benign downstream transformations and advanced adversarial attacks.
Comprehensive experiments demonstrate that our training-free framework provides a practical, provenance-by-design defense infrastructure for enforcing content authenticity and combating digital misattribution in the generative AI era.

\bibliography{references}

\end{document}